\documentclass[useAMS,usenatbib]{mn2e}

\pdfoutput=1

\usepackage{epsf, epsfig}

\usepackage[shortcuts,acronyms]{glossaries-extra}		
\usepackage{enumitem}
\usepackage{amssymb}									
\usepackage{siunitx}									
\usepackage{etoolbox}          						
\usepackage{caption}									
\usepackage{placeins}									

\renewcommand{\bfseries}{\fontseries{b}\selectfont}
\robustify\bfseries             
\newrobustcmd{\B}{\bfseries}    

\def\href#1#2{#2}
\def\urlinner#1{#1\endgroup}
\def\url{\begingroup\def\do##1{\catcode`##1 12 }%
  \do\\\do\$\do\&\do\#\do\^\do\_\do\%\do\~ \ttfamily \urlinner}

\newabbr[longplural = ultra-compact dwarf galaxies]{UCD}{UCD}{ultra-compact dwarf galaxy}
\newabbr{GC}{GC}{globular cluster}
\newabbr{DF}{DF}{dynamical friction}
\newabbr[longplural = centres of mass]{COM}{COM}{centre of mass}
\newabbr{BH}{BH}{black hole}
\newabbr{MBH}{MBH}{massive black hole}
\newabbr{IMBH}{IMBH}{intermediate-mass black hole}
\newabbr[longplural = active galactic nuclei]{AGN}{AGN}{active galactic nucleus}
\newabbr{NSC}{NSC}{nuclear star cluster}
\newabbr{GW}{GW}{gravitational wave}
\newabbr{ISM}{ISM}{interstellar medium}

\expandafter\def\csname model_mn01 \endcsname{M1}
\expandafter\def\csname model_mn02 \endcsname{M2}
\expandafter\def\csname model_mn03 \endcsname{M3}
\expandafter\def\csname model_mn04 \endcsname{M4}
\expandafter\def\csname model_mn05 \endcsname{M5}
\expandafter\def\csname model_mn06 \endcsname{M6}
\expandafter\def\csname model_mn07 \endcsname{M7}
\expandafter\def\csname model_mn08 \endcsname{M8}
\expandafter\def\csname model_mn09 \endcsname{M9}
\expandafter\def\csname model_mn091 \endcsname{M19}
\expandafter\def\csname model_mn092 \endcsname{M20}
\expandafter\def\csname model_mn093 \endcsname{M21}
\expandafter\def\csname model_mn094 \endcsname{M22}
\expandafter\def\csname model_mn095 \endcsname{M23}
\expandafter\def\csname model_mn010 \endcsname{M10}
\expandafter\def\csname model_mn011 \endcsname{M11}
\expandafter\def\csname model_mn012 \endcsname{M12}
\expandafter\def\csname model_mn013 \endcsname{M13}
\expandafter\def\csname model_mn014 \endcsname{M14}
\expandafter\def\csname model_mn015 \endcsname{M15}
\expandafter\def\csname model_mn016 \endcsname{M16}
\expandafter\def\csname model_mn017 \endcsname{M17}
\expandafter\def\csname model_mn018 \endcsname{M18}
\expandafter\def\csname model_imn01 \endcsname{M24}
\expandafter\def\csname model_imn02 \endcsname{M25}
\expandafter\def\csname model_imn03 \endcsname{M26}
\expandafter\def\csname model_imn04 \endcsname{M27}
\expandafter\def\csname model_mn028 \endcsname{M28}
\expandafter\def\csname model_mn101 \endcsname{M29}
\expandafter\def\csname model_mn102 \endcsname{M30}
\expandafter\def\csname model_mn103 \endcsname{M31}
\expandafter\def\csname model_mn104 \endcsname{M32}

\title[Formation of massive black holes]{Formation  of massive black holes 
in ultra-compact dwarf galaxies: migration of primordial intermediate-mass
black holes in N-body simulation} 

\author[H. Wirth and K. Bekki]
{Henriette Wirth${}^1$\thanks{E-mail:
henri-ette\_w@web.de} and
Kenji Bekki${}^2$ \\ 
${}^1$Elektronische Fahrwerksysteme GmbH,
Dr.-Ludwig-Kraus-Str. 6,
85080 Gaimersheim, Germany
\\
${}^2$ICRAR, M468,
The University of Western Australia,
35 Stirling Hwy, Crawley,
Western Australia 6009, Australia
\\}
  
\begin{document}

\date{Accepted, Received 2005 February 20; in original form }

\pagerange{\pageref{firstpage}--\pageref{lastpage}} \pubyear{2005}

\maketitle

\label{firstpage}

\begin{abstract}

Recent observational studies of \acp{UCD} have discovered \acp{MBH}, with masses of more than ${10^6~\rm M_\odot}$, in their central regions.
We here consider that these \acp{MBH} can be formed through merging of \acp{IMBH}, with masses of ${[10^3-10^5]~{\rm M}_{\odot}}$,
within the stellar nuclei of dwarf galaxies, which are progenitors
of \acp{UCD}.
We numerically investigate this formation process for a wide range of model
parameters using N-body simulations.
This means that \ac{IMBH} growth and feedback is neglected in this study.
We find that only massive \acp{IMBH} of $10^5~\rm M_\odot$ sink into the central regions of their host dwarf ($\approx 10^{10}~\rm M_\odot$) to be gravitationally trapped
by its stellar nucleus within less than 1 Gyr in most dwarf models.
We also find that lighter \acp{IMBH} with $[1 - 30] \times 10^3~\rm M_\odot$ sink into the centre in low-mass dwarfs ($\approx 10^{9}~\rm M_\odot$) due to more efficient \ac{DF}.
Additionally, we show that the \acp{IMBH} can form binaries in the centre and, rarely, before they reach the centre, which may lead to the \acp{IMBH} merging and thus emitting gravitational waves that could be detected by LISA.
Finally, we discuss the required number of \acp{IMBH} for the \ac{MBH} formation in \acp{UCD} and the physical roles of stellar nuclei in \ac{IMBH} binaries and mergers.
\end{abstract}


\glsresetall

\section{Introduction}

\Acp{UCD} are very compact stellar systems that were discovered in the Fornax
cluster of galaxies about two decades ago \citep[e.g.][]{1999A&AS..134...75H, 2000PASA...17..227D}.
With their typical sizes of 10 - 100
pc and masses between $10^6$ to ${10^8~\rm M_\odot}$
\citep[e.g.][]{Drinkwater_2003,2005ApJ...627..203H,2008A&A...487..921M}
they link \acp{GC} and dwarf galaxies.
However, they were shown to be distinct from both because of
their lower surface brightness to core luminosity ratio
\citep{Drinkwater_2003}.
Various physical properties of \acp{UCD}, revealed by observation, including internal structural
properties,  star formation histories, and age-metallicity relation, have shown similarities to both \acp{GC} and galaxies for different \acp{UCD} \citep[e.g.][]{Evstigneeva_2007,2008A&A...487..921M,2009MNRAS.394L..97H,2011AJ....142..199B,2015MNRAS.451.3615N}.
As a consequence, it has been suggested that there might be several \ac{UCD} subpopulations of different origins \citep[e.g.][]{2011MNRAS.414..739N,2011AJ....142..199B}.
Although several formation mechanisms have been proposed for UCDs, including
tidal stripping of nucleated dwarf galaxies \citep[e.g.][]{2001ApJ...552L.105B,2019lgei.confE..33M},
merging of  star clusters \citep[e.g.][]{2002MNRAS.330..642F},
and starbursts in super-massive molecular clouds \citep{2018MNRAS.478.3564G}, no theory is yet to explain these observations in a fully self-consistent manner.

The observed mass to luminosity ratio, which in most \acp{UCD} is about
twice as large as that of galactic \acp{GC} \citep{Drinkwater_2003,2011AJ....142..199B}, is an important key observed value that needs to be explained by the formation  scenarios of UCDs.
Different causes like a top-heavy stellar initial mass function
\citep{2009MNRAS.394.1529D}, dark matter \citep{2008MNRAS.391.942B} and central
\acp{MBH} \citep{2013A&A...558A..14M} have been proposed as an explanation for
these enhanced mass to luminosity ratios. The latter is especially interesting
as \acp{MBH} have been observed in \acp{UCD} recently.
Evidence for three
\acp{MBH} has been found for Virgo Cluster \acp{UCD} through dynamic modelling \citep{2014Natur.513..398S,2017ApJ...839...72A}.
Similarly,
\cite{10.1093/mnras/sty913} found evidence for a $3.5 \times {10^6 ~ \rm M_\odot}$
\ac{MBH} in the centre of UCD3 in the Fornax cluster.
From the mass to luminosity ratio of a sample of 49 \acp{UCD} \cite{2013A&A...558A..14M} estimated that about half of the \acp{UCD} could be expected to host a central \ac{BH} with a noticeable effect on observations.

The obvious question to ask now would be: How did those \acp{MBH} get there?
Although UCD formation processes have been investigated in previous theoretical models \citep[e.g.][]{2003MNRAS.344..399B,2013MNRAS.433.1997P,2012A&A...537A...3M,10.1093/mnras/sty913}, the \ac{MBH} formation in \acp{UCD} formation is yet to be fully explored.
The two main formation channels proposed are: \acp{UCD} as the most massive star clusters or nuclei of stripped galaxies \citep[e.g.][]{2012A&A...537A...3M,10.1093/mnras/sty913}.
\cite{2013A&A...558A..14M} proposed that the \acp{MBH} could be inherited from a progenitor galaxy from which the \acp{UCD} could have formed.
However, there are no theoretical studies about the formation of those \acp{MBH} yet.
If they are indeed inherited from a progenitor galaxy, we should have a closer look at the properties of \acp{BH} in dwarf galaxies.

In the centres of many galaxies evidence for the existence of \acp{MBH} was found
\citep[e.g.][]{1995PASP..107..803U,2013ARA&A..51..511K}, many of which coexist with nuclear clusters \citep{2009MNRAS.397.2148G}.
The formation of these
\acp{MBH} is yet to be understood and many different formation scenarios have
been proposed \citep{2010A&ARv..18..279V}. One interesting scenario suggested
by \cite{2003IAUS..208..157E} is that \acp{MBH} could have been formed by
\ac{IMBH} mergers. While plenty of evidence for \acp{MBH} has been detected,
only few promissing \acp{IMBH} candidates have been found
\citep{2017IJMPD..2630021M}.

Similar to the \ac{MBH} formation there are
different hypotheses for the formation of \acp{IMBH}. They could have grown
from stellar mass \acp{BH} due to accretion and mergers \citep{2019MNRAS.487.5549B},
however, a collapse of a massive Population III star
\citep{2014ApJ...781...60H}, the direct collapse of a gas cloud \citep{10.1111/j.1365-2966.2006.10801.x,2014MNRAS.439.1092T} or runaway collisions
in dense metal-poor clusters \citep{2009ApJ...694..302D,2016MNRAS.459.3432M}
are also possible scenarios to form an \ac{IMBH}. A
more detailed overview over different proposed \ac{IMBH} formation scenarios
can be found in \cite{2017IJMPD..2630021M}.
Little is known about the number of \acp{IMBH} in dwarf galaxies.
\cite{10.1111/j.1365-2966.2007.11528.x} derived an upper limit of one \ac{IMBH} in the disk and up to 1000 in the halo of the gas rich dwarf galaxy Holmberg II by comparing simulated X-ray sources to observations.
It is, however, unknown what the limit of \acp{IMBH} in dwarf galaxies is in general.

The purpose of this paper is to investigate whether the \acp{MBH} in the centres of \acp{UCD} could have been build initially from \acp{IMBH} within the nucleated dwarf galaxies, from which UCDs can originate.
Using numerical simulations of nucleated dwarf galaxies with \acp{IMBH}, we particularly investigate the dynamical evolution of \acp{IMBH} in their host dwarfs.
We try to find out whether the effect of \ac{DF} on our \acp{IMBH} is strong
enough for them to spiral into the nucleus before the galaxy is tidally
stripped. We also investigate whether they can be trapped there and form a central cluster in
which they could merge to form an \ac{MBH}.
Although previous observational and theoretical studies discussed the physical relationships between stellar galactic nuclei, \acp{MBH}, and their host galaxies \citep[e.g.][]{2010ApJ...714L.313B,2015ApJ...812...72A,2016MNRAS.457.2122G,2017MNRAS.472.4013C,2019MNRAS.484..520A}, we focus exclusively on the formation of \acp{MBH} in \acp{UCD} (i.e. stellar nuclei).

The plan of this paper is as follows:
In section \ref{sec_Model} we will describe the model used.
We give a short overview over the code used in section \ref{sec_Code}.
The details of the simulated galaxies and \acp{IMBH} are explained in \ref{sec_Galaxy} and \ref{sec_IMBH} respectively and we describe the main parameters used in section \ref{sec_Param}.
We show our results in section \ref{sec_Results}.
We discuss the efficiency of \ac{DF} in section \ref{sec_DF}, what fraction of \acp{IMBH} was trapped in the nucleus in section
\ref{sec_NucTrap} and the binaries we found, that potentially could lead to
mergers, in section \ref{sec_Bin}.
We discuss whether \ac{IMBH} mergers alone are sufficient to explain the formation of \acp{MBH}, the potential formation of \acp{GW} and future work in sections \ref{sec_Disc_enoughBH}, \ref{sec_Disc_GW} and \ref{sec_Disc_FW} respectively.
Finally we present our conclusions in section \ref{sec_conc}.

\begin{table}
\centering
\begin{minipage}{80mm}
\caption{Description of the basic parameter values
for the fiducial galaxy  model.}
\label{table_fiducialMod}
\begin{tabular}{lr}
\hline
Physical properties
&  Values  \\\hline
DM mass  & $1.0 \times 10^{10} ~ {\rm M}_{\odot}$  \\
DM profile  & NFW  \\
Virial radius & 12.3 kpc \\
{$c$  \footnote{$c$ is the $c$-parameter in the NFW dark matter
profiles.}}
&  16 \\
Dwarf morphology  &  disky   \\
Stellar disk mass & $3.6 \times 10^{8} ~ {\rm M}_{\odot}$ \\
Gas disk  mass & $0.0 \times 10^{10} ~ {\rm M}_{\odot}$ \\
Stellar nucleus  & $1.8 \times 10^7 ~ {\rm M}_{\odot}$ \\
MBH number  &  10  \\
MBH  mass & $1.0 \times 10^{5} ~ {\rm M}_{\odot}$ \\
MBH growth &  Not included \\
MBH-ISM interaction  &  Not included \\
{$\epsilon$ (DM)  \footnote{$\epsilon$ is the
gravitational softening length for particles.}}
&   131.9 pc \\ 
$\epsilon$ (stars)  &  5.9 pc  \\ 
$\epsilon$ (nucleus) & 0.5 pc \\ 
Mass resolution (DM)  &  $5 \times 10^4 ~ {\rm M}_{\odot}$  \\ 
Mass resolution (stars)  &  $360 ~ {\rm M}_{\odot}$  \\ 
Mass resolution (nucleus)  &  $180 ~ {\rm M}_{\odot}$  \\ 
Time step width & ${1.41 \times 10^5 ~ {\rm yr}}$\\\hline
\end{tabular}
\end{minipage}
\end{table}

\begin{table*}
\centering
\begin{minipage}{180mm}
\caption{The model parameters for dwarfs and IMBHs in simulations.
$m_{\rm p}$ is the mass of a disk particle.}
\label{table_ModelParam}
\begin{tabular}{l
				*{7}{S[detect-weight,
           			mode=text,       
           			table-format=1.3]}
           		l}
\hline
ID
& {$M_{\rm dm}$ ($10^{10} {\rm M}_{\odot}$)}
& {$M_{\rm s}$ ($10^{8} {\rm M}_{\odot}$)}
& {$R_{\rm s}$ (kpc)}
& {$M_{\rm bh}$ ($10^{5} {\rm M}_{\odot}$)}
& {$\frac{M_{\rm bh}}{m_{\rm p}}$}
& {$R_{\rm bh}$ (pc)}
& {$N_{\rm bh}$}
& Comments \\ 
\hline
\csname model_mn01 \endcsname & 1.0 & 3.6 & 0.88 & 0.3 & 83.3 & 176 & 10 & fiducial model \\
\csname model_mn02 \endcsname & 1.0 & 3.6 & 0.88 & 0.1 & 27.8 & 176 & 10 & \\
\csname model_mn03 \endcsname & 1.0 & 3.6 & 0.88 & 1.0 & 277.8 & 176 & 10 & \\
\csname model_mn04 \endcsname & 1.0 & 3.6 & 0.88 & 0.03 & 8.3 & 176 & 10 & \\
\csname model_mn05 \endcsname & 1.0 & 3.6 & 0.88 & 0.3 & 83.3 & 176 & 10 & no nucleus \\
\csname model_mn06 \endcsname & 1.0 & 3.6 & 0.88 & 0.1 & 27.8 & 176 & 10 & no nucleus \\
\csname model_mn07 \endcsname & 1.0 & 3.6 & 0.88 & 1.0 & 277.8 & 176 & 10 & no nucleus \\
\csname model_mn08 \endcsname & 1.0 & 3.6 & 0.88 & 0.03 & 8.3 & 176 & 10 & no nucleus \\
\csname model_mn09 \endcsname & 0.1 & 0.36 & 0.55 & 0.3 & 833.3 & 110 & 10 & low-mass model \\
\csname model_mn010 \endcsname & 0.1 & 0.36 & 0.55 & 0.1 & 277.8 & 110 & 10 & low-mass model \\
\csname model_mn011 \endcsname & 0.1 & 0.36 & 0.55 & 1.0 & 2777.8 & 110 & 10 & low-mass model \\
\csname model_mn012 \endcsname & 0.1 & 0.36 & 0.55 & 0.03 & 83.3 & 110 & 10 & low-mass model \\
\csname model_mn013 \endcsname & 0.1 & 0.36 & 0.55 & 0.3 & 833.3 & 275 & 10 & low-mass model \\
\csname model_mn014 \endcsname & 0.1 & 0.36 & 0.55 & 0.1 & 277.8 & 275 & 10 & low-mass model \\
\csname model_mn015 \endcsname & 0.1 & 0.36 & 0.55 & 1.0 & 2777.8 & 275 & 10 & low-mass model \\
\csname model_mn016 \endcsname & 0.1 & 0.36 & 0.55 & 0.03 & 83.3 & 275 & 10 & low-mass model \\
\csname model_mn017 \endcsname & 1.0 & 3.6 & 0.88 & 0.3 & 83.3 & {[44 - 440]} & 10 & distance varies linearly \\
\csname model_mn018 \endcsname & 1.0 & 3.6 & 0.88 & 1.0 & 833.3 & {[44 - 440]} & 10 & distance varies linearly \\
\csname model_mn091 \endcsname & 1.0 & 3.6 & 1.75 & 0.3 & 83.3 & 350 & 10 &  \\
\csname model_mn092 \endcsname & 0.3 & 1.3 & 0.48 & 0.1 & 92.6 & 96 & 10 &  \\
\csname model_mn093 \endcsname & 1.0 & 3.6 & 0.88 & {[0.01 - 3]} & {[2.8 - 833.3]} & 176 & 6 & different $M_{\rm bh}$ masses \\
\csname model_mn094 \endcsname & 1.0 & 3.6 & 0.88 & 0.3 & 83.3 & 176 & 10 &  \\
\csname model_mn095 \endcsname & 1.0 & 3.6 & 0.88 & {[0.01 - 3]} & {[2.8 - 833.3]} & 176 & 6 & different $M_{\rm bh}$ masses \\
\csname model_imn01 \endcsname & 1.0 & 3.6 & 0.88 & 0.1 & 27.8 & {[88 - 880]} & 10 & distance varies linearly \\
\csname model_imn02 \endcsname & 1.0 & 3.6 & 0.88 & 0.3 & 83.3 & {[88 - 880]} & 10 & distance varies linearly \\
\csname model_imn03 \endcsname & 1.0 & 3.6 & 0.88 & 1.0 & 277.8 & {[88 - 880]} & 10 & distance varies linearly \\
\csname model_imn04 \endcsname & 1.0 & 3.6 & 0.88 & 0.03 & 8.3 & {[88 - 880]} & 10 & distance varies linearly \\
\csname model_mn028 \endcsname & 1.0 & 3.6 & 0.88 & 10.0 & 2777.8 & {[88 - 528]} & 6 & distance varies linearly \\
\csname model_mn101 \endcsname & 3.0 & 10.8 & 1.52 & 1.0 & 92.6 & 305 & 10 & \\
\csname model_mn102 \endcsname & 0.03 & 0.1 & 0.15 & 0.1 & 925.9 & 31 & 10 & \\
\csname model_mn103 \endcsname & 0.03 & 0.1 & 0.15 & 0.01 & 92.6 & 31 & 10 & \\
\csname model_mn104 \endcsname & 3.0 & 10.8 & 1.52 & 10.0 & 925.9 & 305 & 10 & \\
\hline
\end{tabular}
\end{minipage}
\end{table*}

\section{The model}
\label{sec_Model}

\subsection{Simulation code for IMBH evolution}
\label{sec_Code}

In order to investigate both (i) the dynamical evolution
of dwarf galaxies and (ii) the orbital evolution of \acp{IMBH} within their host dwarfs
in a self-consistent manner, we adopt our code for direct Nbody simulations
used for the evolution of \acp{GC} in dwarfs \citep[BT16]{2016ApJ...831...70B}. Since the details of the code
are given in BT16, we here briefly describe the code.
The gravitational softening length ($\epsilon$) can be chosen separately
for each of the components in a galaxy (e.g., halo, disk, and nucleus) for 
the adopted numbers of particles of the components. The maximum timestep width
($\delta t$) is chosen to be rather small (${[10^4-10^5]~\rm yr}$) in comparison with
galaxy-scale simulations, though such narrow $\delta t$  is not required
for the dynamical evolution of dwarf galaxies.
We did not include gas dynamics, star formation, chemical evolution, stellar feedback effects, the formation and evolution of dust and molecular gas formation on dust grains in the present study, though our other galaxy-scale simulations included these processes in a self-consistent manner \citep{2007PASA...24...77B,2013MNRAS.432.2298B}.
This is mainly because because orbital evolution of \acp{IMBH} might not be influenced by such baryonic processes.
However, if efficient accretion of cold gas onto IMBHs is possible in dwarfs, then the orbital evolution of IMBHs can be significantly influenced by such a process.
We will discuss how this IMBH growth via gas accretion can influence MBH formation in our future works.

\subsection{Host galaxies for IMBHs}
\label{sec_Galaxy}

We assume that the host galaxy for \acp{IMBH} is a dwarf disk galaxy with stellar galactic nucleus
embedded in a massive dark matter halo. 
In this preliminary works, we only investigate the models with no gas,
though hydrodynamical interaction between ISM and \acp{IMBH} can possibly influence
the orbital evolution of \acp{IMBH} within dwarf galaxies.
The total masses of dark matter halo, stellar disk,  and
nucleus of a dwarf galaxy are denoted as $M_{\rm dm}$, $M_{\rm s}$, and $M_{\rm nuc}$,
respectively.
We adopt the density distribution of the ``NFW''
halo \citep{1996ApJ...462..563N} derived from previous CDM simulations
in order to describe the initial density profile of dark matter halo
in a dwarf galaxy with \acp{IMBH}:
\begin{equation}
{\rho}(r)=\frac{\rho_{0}}{(r/r_{\rm s})(1+r/r_{\rm s})^2},
\end{equation}
where  $r$, $\rho_{0}$, and $r_{\rm s}$ are
the spherical radius,  the characteristic  density of a dark halo,  and the
scale
length of the halo, respectively.
The $c$-parameter ($c=r_{\rm vir}/r_{\rm s}$, where $r_{\rm vir}$ is the virial
radius of a dark matter halo) and $r_{\rm vir}$ are chosen appropriately
for a given dark halo mass ($M_{\rm dm}$)
by using the $c-M_{\rm h}$ relation for $z=0$
predicted by recent cosmological simulations
\citep[e.g.][]{2007MNRAS.381.1450N}. For the adopted mass ranges of $M_{\rm dm}$,
we consider that $c=16$ is a quite reasonable value.
In the present study, we mainly investigate dwarf galaxies with 
$M_{\rm dm}$ ranging from
${10^{9}~{\rm M}_{\odot}}$ to
${10^{10}~{\rm M}_{\odot}}$
and $R_{\rm vir}$ ranging from 7.7 kpc to 24.5 kpc.

We assume that the stellar disk of a dwarf galaxy can be represented
by the so-called exponential profile. Accordingly,  
the radial ($R$) and vertical ($Z$) density profiles of the stellar disk are
assumed to be proportional to $\exp (-R/R_{0}) $ with scale
length $R_{0} = 0.2R_{\rm s}$  and to ${\rm sech}^2 (Z/Z_{0})$ with scale
length $Z_{0} = 0.04R_{\rm s}$, respectively,
where $R_{\rm s}$ is the size of the stellar disk.
In addition to the
rotational velocity caused by the gravitational field of disk,
bulge, and dark halo components, the initial radial and azimuthal
velocity dispersions are assigned to the disc component according to
the epicyclic theory with Toomre's parameter $Q$ = 1.5.
The vertical velocity dispersion at a given radius is set to be 0.5
times as large as the radial velocity dispersion at that point.
The mass-ratio of the stellar disk to its dark matter halo
is assumed to rather small ranging from 0.01 to 0.03, which is consistent
with the mass scaling relation between stars and dark matter observed in low-mass
galaxies \citep[e.g.][]{2012ApJ...759..138P}.
It could be possible that small disk galaxies can have small bulges, we do not investigate
the models with small bulges in the present study. Small bulges are highly unlikely
to influence the orbital evolution of \acp{IMBH}.

We assume that an initial stellar nucleus in a dwarf galaxy
has a Plummer spherical density profile
\citep[e.g.][]{1987gady.book.....B} 
with a total stellar mass ($M_{\rm nuc}$),
and a size ($R_{\rm nuc}$).
In a Plummer model,
the scale length ($a_{\rm nuc}$) of the system is determined by the formula
\begin{equation}
a_{\rm nuc} = GM_{\rm nuc}/6{{\sigma}_{\rm nuc}}^{2}, \;
\end{equation}
where G is the gravitational constant and
${\sigma}_{\rm nuc}$ is
a central
velocity dispersion of the nucleus.
In the present study, we do not consider initial angular momentum
of the nucleus, because the orbital evolution of \acp{IMBH} might not be 
influenced by the angular momentum.
Therefore,
the above equation is appropriate for the adopted stellar
systems with no initial angular
momentum (i.e., dynamically supported only by velocity dispersion).
Since observational studies showed that the total masses of stellar galactic nuclei
can be proportional to those of the stellar components of their host galaxies
(e.g., Cote et al. 2006), we assume that $M_{\rm nuc}$ is proportional to
$M_{\rm s}$.
We adopt $M_{\rm nuc}=1.8 \times {10^7~{\rm M}_{\odot}}$ 
and $R_{\rm nuc}=44$ pc
for $M_{\rm dm}={10^{10}~{\rm M}_{\odot}}$
and $M_{\rm nuc}=1.8 \times {10^6~{\rm M}_{\odot}}$ 
and $R_{\rm nuc}=28$ pc
for $M_{\rm dm}={10^{9}~{\rm M}_{\odot}}$.

\subsection{IMBH models}
\label{sec_IMBH}

Each \ac{IMBH} in a dwarf galaxy is represented by a point-mass particle with
a mass ($M_{\rm bh}$), and it is initially in the host galaxy's disk: \acp{IMBH} initially
in the dark matter halo are not considered in the present study,
though they
can possibly sink into the central region due to \ac{DF} against
the dark matter particles.
Each \ac{IMBH} is assumed to have a circular velocity at its initial position (i.e.,
no radial motion initially) and different \acp{IMBH} have different initial positions
within their dwarf galaxy. Although recent cosmological simulations have investigated
the evolution of 3D positions of primordial \acp{IMBH} \citep[e.g.][]{2019MNRAS.487.5549B}, there 
are no robust predictions provided for the 3D positions
of IMBHs within their
host dwarf galaxies. We thus investigate a larger number
of models with different \ac{IMBH}
positions in the present study.
We assume that each dwarf galaxy can contain 10 \acp{IMBH} within its disk, though
it is not theoretically and observationally clear how many \acp{IMBH} can possibly
exist in a dwarf galaxy.

We consider that the following parameter, $R_{\rm m}$,
is quite important in the orbital evolution of \acp{IMBH} due to \ac{DF} in dwarfs:
\begin{equation}
R_{\rm m} = \frac{ m_{\rm p} }{ M_{\rm bh} },
\end{equation}
where $m_{\rm p}$ is the mass of a  particle.
If this mass-ratio $R_{\rm m}$ is quite small, then
dynamical friction of \acp{IMBH} can be properly investigated.
For example, $R_{\rm m}$ for disk star particles is $<0.1$ in the
models with  $M_{\rm bh} \ge
3 \times 10^3 {\rm M}_{\odot}$. Therefore,
we can properly investigate the \ac{DF} of
\acp{IMBH} against disk field stars in dwarfs.  However, $R_{\rm m}$
for dark matter particles can be  larger than 1 in the models
with low-mass \acp{IMBH}. Accordingly, we cannot investigate the dynamical
friction of \acp{IMBH} against dark matter in the present study.

\subsection{Parameter study}
\label{sec_Param}

We consider that the model \csname model_mn01 \endcsname\ is  the fiducial model in which
10 \acp{IMBH} are assumed to be moving within a dwarf galaxy
with $M_{\rm dm}={10^{10}~{\rm M}_{\odot}}$,
$M_{\rm s}=3.6 \times {10^{8}~{\rm M}_{\odot}}$
and $R_{\rm s}=875$ pc, because this model and those
with different \ac{IMBH} masses shows interesting behaviours
of orbital evolution of \acp{IMBH}.
The total number of particles in the fiducial model  is
200000, 1000000, and 100000 for dark matter, stellar disk, and stellar nucleus, which leads to particle masses of $50000$, $360$ and $180~\rm M_\odot$ respectively.
These large particle masses are required to keep the computational cost reasonably low, however, as we can see in Table \ref{table_ModelParam} the ratios between the \ac{IMBH} masses and the stellar masses are still large enough to realistically simulate the effect dynamical friction would have on the \acp{IMBH}.
Different softening lengths are allocated for different components
of a dwarf (dark matter, stellar disk, and nucleus) so that the evolution
of \acp{IMBH} within the stellar disk and its nucleus can be both investigated
self-consistently.
The mass and size resolution in the model are
${360~{\rm M}_{\odot}}$ and 5.9 pc, respectively, for the stellar disk.
The \acp{IMBH} are treated as massive disk particles and therefore have the same softening length as the disk stars (${5.9~\rm pc}$).
The parametres of the fiducial model are summarized in Table \ref{table_fiducialMod}.
We mainly investigate 27 models with different model parameters  and the parameter
values and ranges are summarized in Table \ref{table_ModelParam}.

To determine whether a central \ac{MBH} can form via merging of several \acp{IMBH} or not we have to investigate three questions, which we will answer in the next sections:
\begin{enumerate}[leftmargin=\parindent,align=left,labelwidth=\parindent,labelsep=0pt,label=\arabic*.]
\item Can \acp{IMBH} spiral into the galaxies centre due to \ac{DF} before the nucleus is stripped by tidal forces?
\item Are enough \acp{IMBH} trapped in the nucleus to form a central cluster of \acp{IMBH}?
\item Do enough \acp{IMBH} merge to form a central \ac{MBH}?
\end{enumerate}

Regarding  the first question, the
\acp{IMBH} need to spiral into the central regions of stellar nuclei
due to dynamical fraction
in dwarfs before the disintegration of dwarfs through tidal interaction
of the dwarfs with their environments (e.g., luminous galaxies, groups,
and clusters). Therefore, the dynamical friction time scale ($t_{\rm df}$)
should be shorter than the disintegration time scale ($t_{\rm dis}$):
\begin{equation}
t_{\rm df} < t_{\rm dis}.
\end{equation}
Previous numerical simulations of UCD formation through galaxy threshing
demonstrated that $t_{\rm dis}$ should be at least a few Gyr
\citep[e.g.][]{2003MNRAS.344..399B}. Therefore, we consider that $t_{\rm df}$ should
be as short as 1 Gyr for \ac{MBH} formation via \ac{IMBH} migration
into stellar nuclei (before host disintegration) in the present study.
Since $t_{\rm dis}$ can be quite different depending on the orbits
of dwarfs within their host environments,
the adopted $t_{\rm df} \approx 1$ Gyr can be a bit too short.

\section{Results}
\label{sec_Results}

\begin{figure}
	\includegraphics[width=\columnwidth]{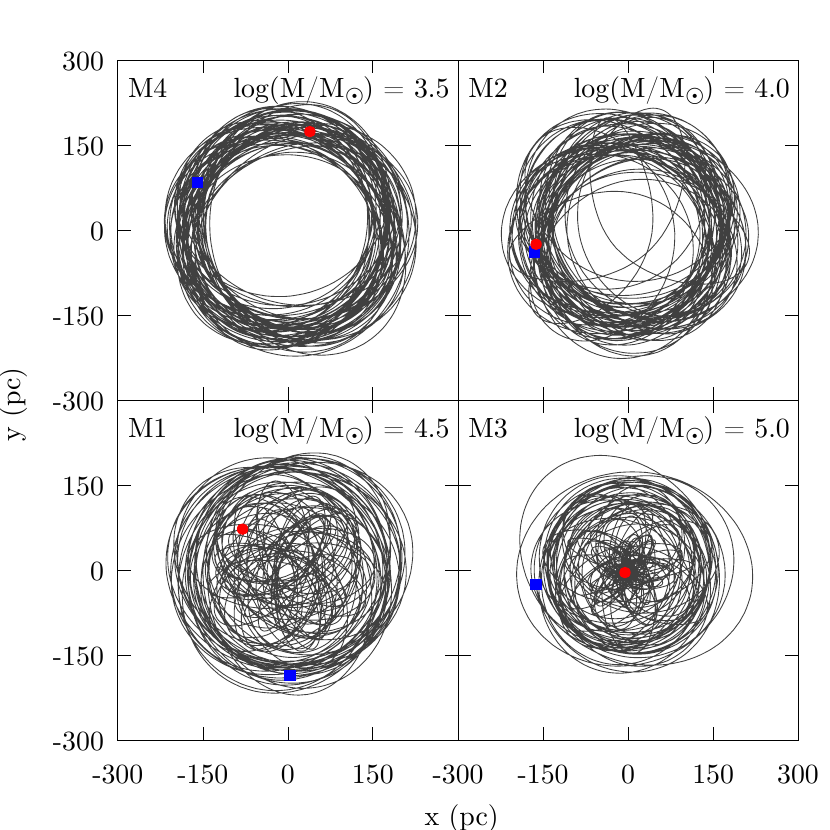}
	\caption{Orbital evolution of IMBHs with different masses over 1 Gyr in four models with $M_{\rm s} = {3.6 \times 10^{10}~\rm M_\odot}$.
	The mass of each IMBH is shown in the top right corner, while the model ID is displayed in the top left corner of each panel.
	The galaxy parameters are the same as for the fiducial model.
	The blue square marks the starting point for each IMBH while the red dot marks its final point.
	}
	\label{fig_ExampleOrbits}
\end{figure}

\begin{figure*}
	\includegraphics{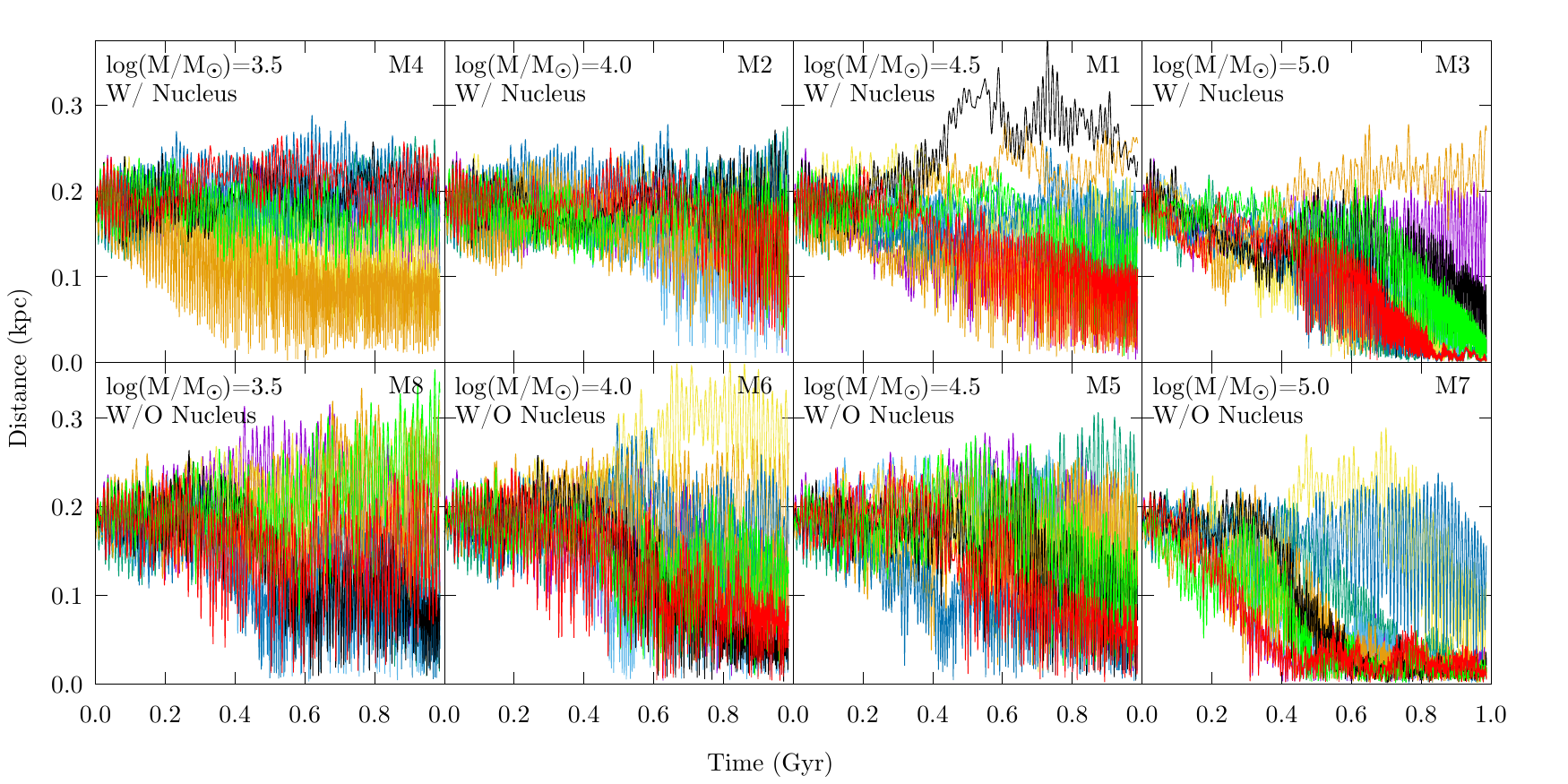}
	\caption{Time evolution of the distances between the IMBHs and the galactic centre.
	The mass of the IMBHs is shown in the upper left.
	W/ Nucleus and W/O Nucleus means the models with or without stellar nuclei, respectively.
	The model names are shown in the top right corner.
	Different colours denote different IMBHs.}
	\label{fig_Distance1}
\end{figure*}

\begin{figure*}
	\includegraphics{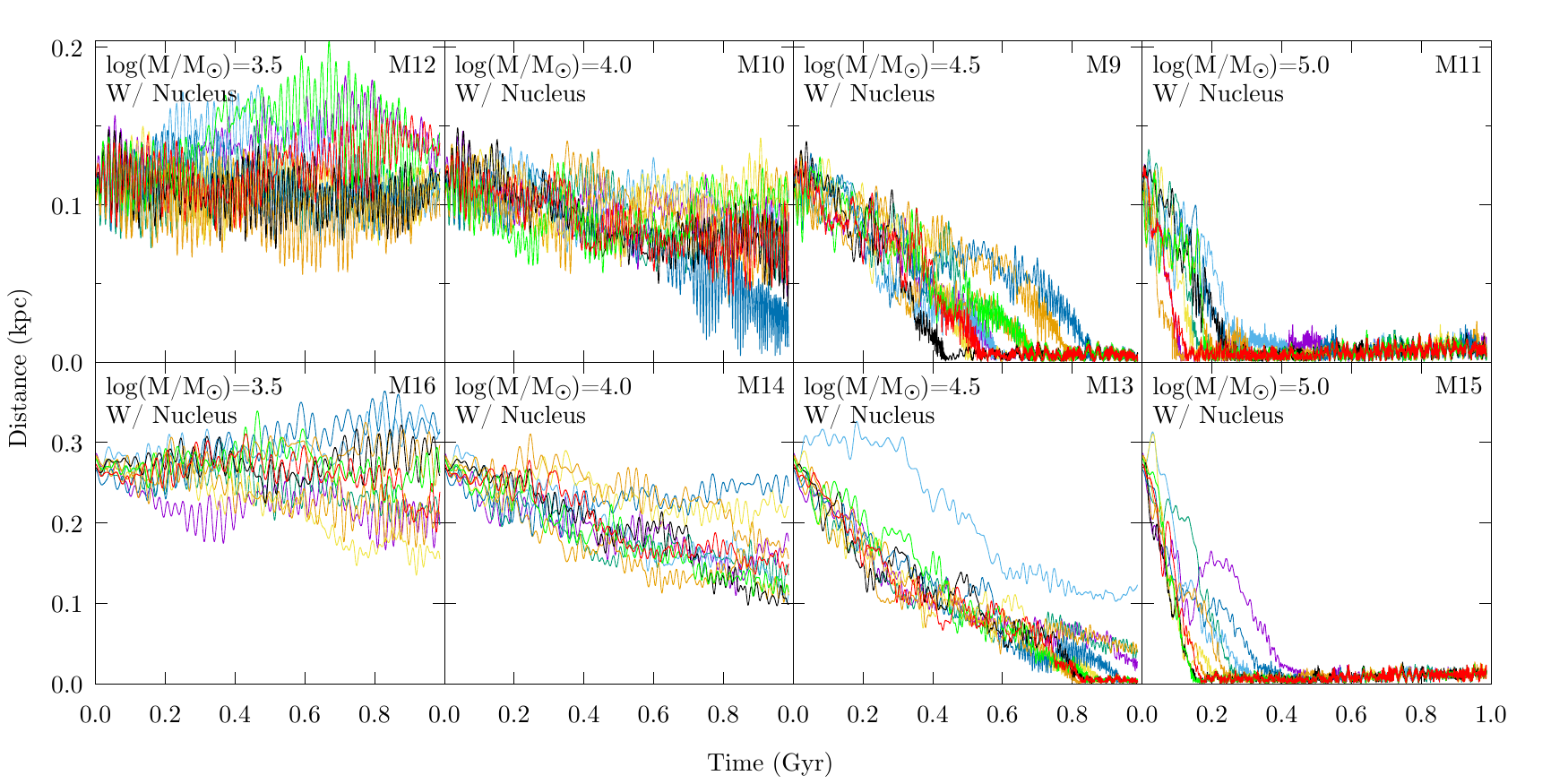}
	\caption{The same as Fig. \ref{fig_Distance1} but for low-mass models with $M_{\rm s} = {3.6 \times 10^7~\rm M_\odot}$. The IMBHs in the top row start at a distance of ${0.2~\rm R_{\rm s}}$ from the centre and the ones in the bottom row from ${0.5~\rm R_{\rm s}}$.}
	\label{fig_Distance2}
\end{figure*}

\begin{figure}
	\includegraphics{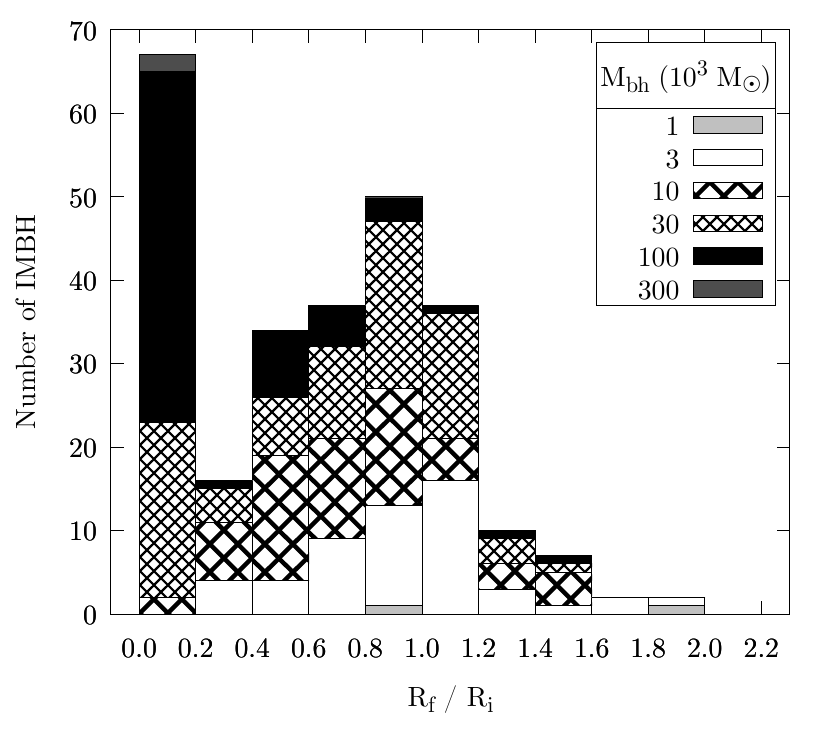}
	\caption{The final distances $R_{\rm f}$ of the IMBH over their initial distances $R_{\rm i}$ are shown for all models. IMBHs of different mass are shown in different colours as shown in the top right.
	The extreme models starting from \csname model_mn028 \endcsname\ are excluded from this graph.}
	\label{fig_RelDist}
\end{figure}

\begin{figure}
	\includegraphics{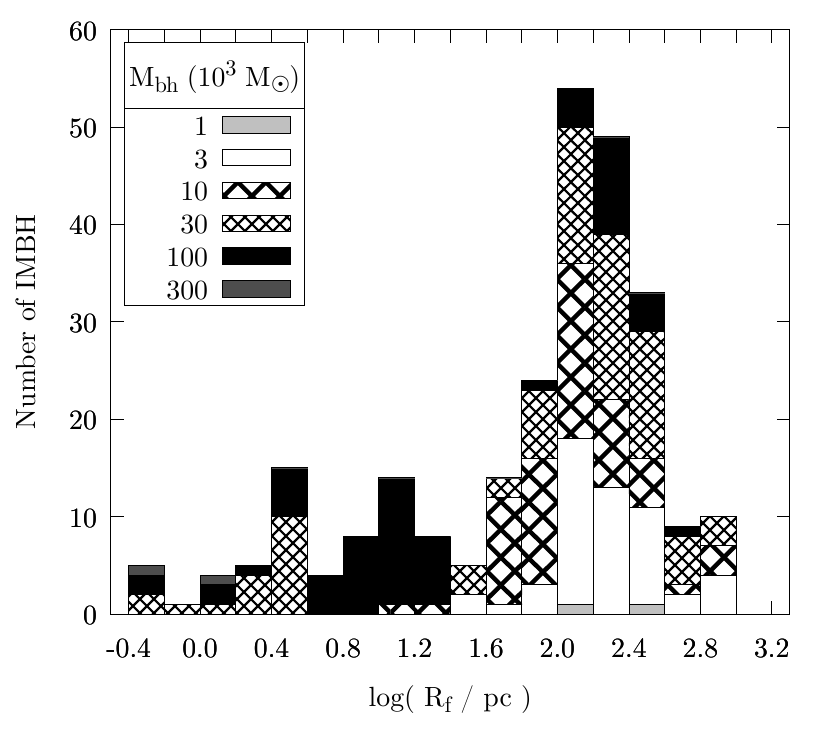}
	\caption{The same as Fig. \ref{fig_RelDist} but for $R_{\rm f}$ (logarithmic scale).}
	\label{fig_LogDist}
\end{figure}

\begin{figure}
	\includegraphics{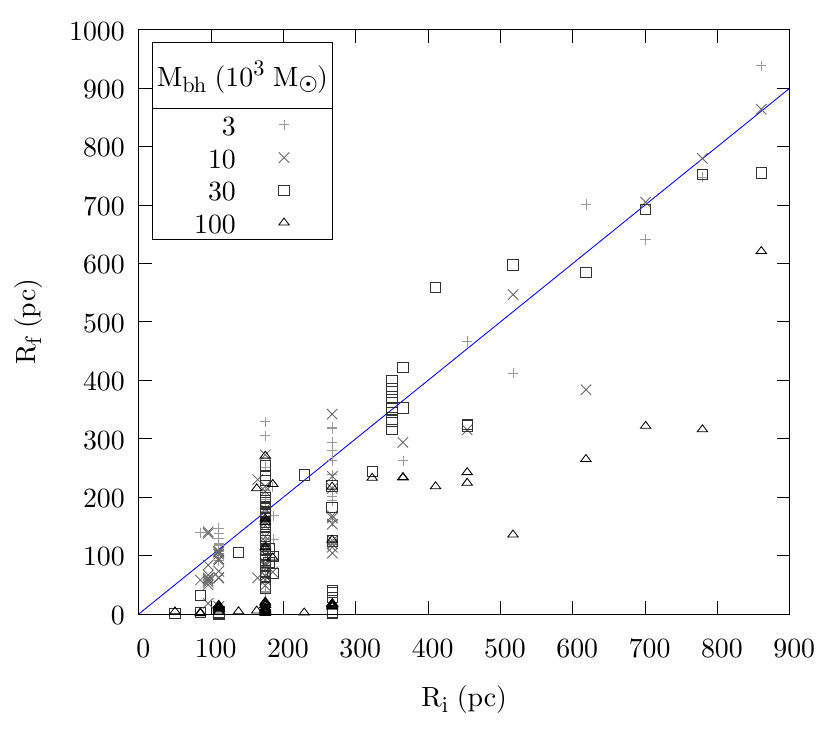}
	\caption{The final distances of the IMBHs in relation to their initial distances for all models. Sorted by the masses of the IMBHs. The identity is shown in blue.
	The extreme models starting from \csname model_mn028 \endcsname\ are excluded from this graph.}
	\label{fig_DistOverIni}
\end{figure}

\subsection{\Ac{DF} depending on the model parameters}
\label{sec_DF}

We start by investigating whether and how \acp{IMBH} can spiral into the nucleus due to dynamical friction within 1 Gyr.
Examples of orbits from our simulations in the fiducial model can be seen in Fig. \ref{fig_ExampleOrbits}.
It is clear that the effect of \ac{DF} against the field stars gets stronger with the mass of the \ac{IMBH} until the \ac{IMBH} ends up near the \ac{COM} of the galaxy, as can be seen in the bottom right panel.
According to theoretical calculations by \cite{1943ApJ....97..255C} the time scale of \ac{DF} is inversely proportional to the \ac{IMBH} mass:
our results are consistent with these analytical predictions.
As we will see below, there could be a threshold mass for the \acp{IMBH} to be sunk into the nucleus within ${1~\rm Gyr}$ within dwarf disk galaxies.

We compiled the results of simulations with different model parameters in Figs. \ref{fig_Distance1} and \ref{fig_Distance2}.
In both Figs. the models are sorted by $M_{\rm bh}$ with the lowest mass ($3 \times {10^3~\rm M_\odot}$) to the far left and the models with the highest masses (${10^5~\rm M_\odot}$) to the far right.
It is clear that only massive \acp{IMBH} can reach the centre within 1 Gyr.
The models in Fig. \ref{fig_Distance1} all have the same disc parameters as our fiducial model.
As we can see, independent of the presence of a nucleus only the \acp{IMBH} with a mass of ${10^5~\rm M_\odot}$ sink into the galaxy's centre within ${1~\rm Gyr}$.
However, \ac{DF} appears to be a little more efficient in models without a nucleus initially.
This can be explained by the lower relative velocities between the \acp{IMBH} and the field stars of the dwarf disk: the \ac{DF} time scale depends also on relative velocities.

Comparing these results to the low-mass models shown in Fig. \ref{fig_Distance2}, we note that for lighter galaxies \acp{IMBH} with ${3 \times 10^4~\rm M_\odot}$ sink towards the centre within ${1~\rm Gyr}$ as well as the ones with ${10^5~\rm M_\odot}$.
The threshold mass is therefore smaller than for our fiducial model.
This lower required mass was to be expected due to the lower velocity dispersion (lower relative velocities between \acp{IMBH} and field stars) in these models. We see a continuation of this trend in appendix \ref{App_ExtGal}, where we discuss \ac{DF} in extreme dwarfs.
Fig. \ref{fig_Distance2} also shows a comparison between models with different initial \ac{IMBH} distances (upper row vs. lower row).
As expected, \acp{IMBH} that start further away from the galactic centre take longer to reach the centre.
However, as we can see, 1 Gyr is still sufficient for \acp{IMBH} with a mass of $3 \times {10^4~\rm M_\odot}$ even if they start at a distance of ${0.5 R_{\rm s}}$ to reach the \ac{COM}.
Comparing between distance from galaxy center and \ac{BH} mass, the latter plays, therefore, a dominant role.

In Fig. \ref{fig_RelDist} we can see the number of \acp{IMBH} depending on their final distance to the \ac{COM} of the nucleus (stars for models without a nucleus) over their initial distance.
As we can see, the majority of the \acp{IMBH} move closer to the \ac{COM}, with a lot of them ending up within 20 per cent of $R_{\rm i}$, where we can find the largest number of \acp{IMBH}.
Again we see that the effect of dynamical friction is stronger for heavy \acp{IMBH}.
No \ac{IMBH} with a mass of less than ${10^4~\rm M_\odot}$ is found within $20$ per cent of its initial radius while most of the \acp{IMBH} with ${10^5~\rm M_\odot}$ can be found in this bin.
As the lighter \acp{IMBH} are not affected by dynamical friction as strongly, a second maximum can be seen around the initial distance.
This shows that the dynamical friction is very weak for those lighter \acp{IMBH}.

The number of \acp{IMBH} depending on the absolute value of their final distance to the \ac{COM} is shown in Fig. \ref{fig_LogDist}.
Again we see that only heavy \acp{IMBH} move towards the \ac{COM}.
Only \acp{IMBH} with masses higher or equal than $3 \times {10^4~\rm M_\odot}$ can be found within 10 pc of the \ac{COM}; and only \acp{IMBH} with masses higher or equal to ${10^4~\rm M_\odot}$ are visible within 20 pc of the \ac{COM}, where we can find 63 \acp{IMBH} (24 per cent of the total \acp{IMBH}).
Only two of those have a mass of ${10^4~\rm M_\odot}$, while the masses of the others are at least ${3 \times 10^4~\rm M_\odot}$.
A maximum can be found between 100 and 158 pc which is close to the initial distances of most of our models (176 pc for our fiducial model, 110 and 225 pc for the low-mass models).
No clear trend is visible within the lowest ${20~\rm pc}$.
This is the region where the \acp{IMBH} form a cluster and therefore influence each other.
Because of our model's limitations (large softening length), we cannot simulate the behaviour of dense \ac{IMBH} clusters accurately.
Future simulations with e.g. NBODY6 will be required to accurately model the behaviour of the \acp{IMBH} in this region.

A comparison between the initial and final distances of the \acp{IMBH} from the \ac{COM} can be seen in Fig. \ref{fig_DistOverIni}.
The identity is shown as a blue line. The graph shows a lot of dispersion due to the chaotic nature of our model.
However, it can be seen that in general the distances are reduced, which is to be expected if the \acp{IMBH} are subject to dynamic friction.
We also see that the effect is strongest for the models with \ac{IMBH} masses of ${10^5~\rm M_\odot}$.
While those heavy \acp{IMBH} can almost all be found far below the identity, showing that they moved closer to the \ac{COM} of the galaxy, the distances of most of the light \acp{IMBH} did not change significantly.

\begin{figure}
	\includegraphics{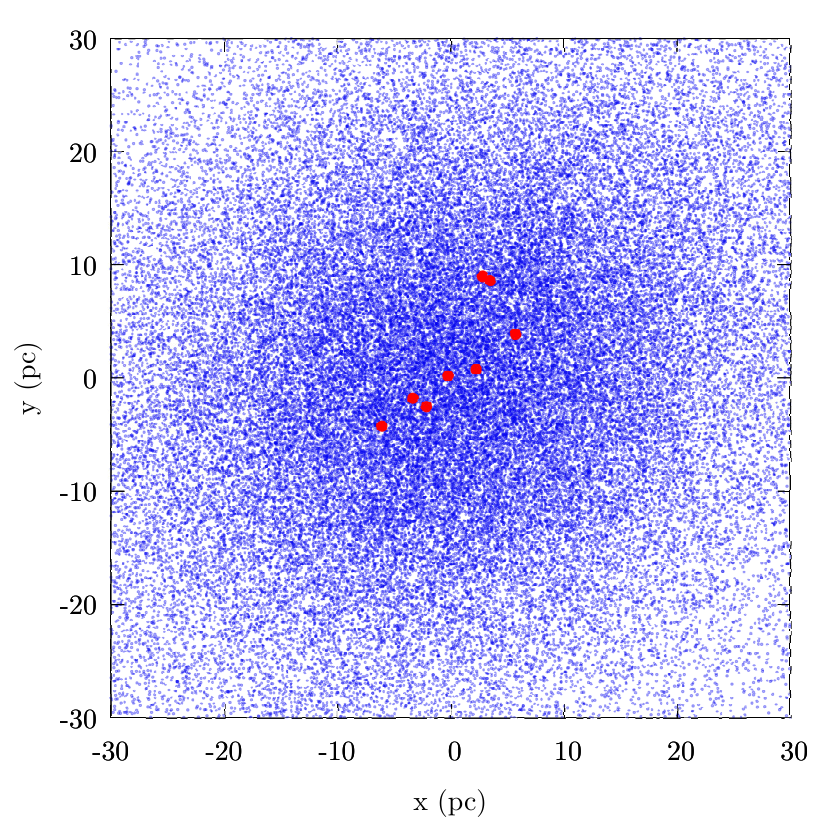}
	\caption{Distribution of stars (blue dots) and IMBHs (big red dots) projected onto the x-y-plane for the central 30 pc of a dwarf in model \csname model_mn03 \endcsname\ after ${1~{\rm Gyr}}$.
	The initial distance of these IMBHs was ${167~\rm pc}$.
	This newly developed IMBH cluster can possibly merge to form a single MBH in the present scenario.}
	\label{fig_Core}
\end{figure}

\begin{figure}
	\includegraphics{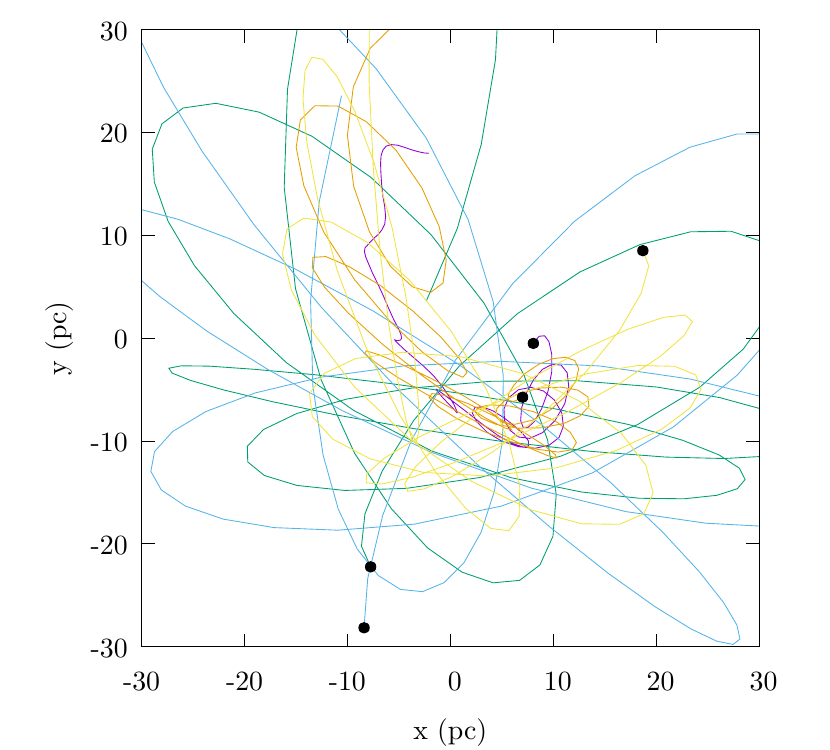}
	\caption{Orbital evolution of IMBHs in model \csname model_mn03 \endcsname.
	The orbits are shown between $T={0.816~\rm Gyr}$ and $T={0.832~\rm Gyr}$. The IMBH positions at $T={0.832~\rm Gyr}$ are marked with black dots.}
	\label{fig_centCluster}
\end{figure}

\subsection{Trapping \acp{IMBH} in the nucleus}
\label{sec_NucTrap}

\begin{table}
\centering
\begin{minipage}{\columnwidth}
\caption{The number of IMBH that are closer than 20 pc or between 20 and 50 pc to the COM of the nucleus for each model. For the models without a nucleus the COM of the galaxy's stars is used instead.}
\label{table_FinalDist}
\begin{tabular}{lrr}
\hline
Model ID 
& $N_{\rm bh}$ ($R_{\rm f} < {20~\rm pc}$)
& $N_{\rm bh}$ (${20~{\rm pc}} < R_{\rm f} < {50~{\rm pc}}$) \\
\hline
\csname model_mn01 \endcsname & 0 & 0 \\
\csname model_mn02 \endcsname & 0 & 0 \\
\csname model_mn03 \endcsname & 8 & 0 \\
\csname model_mn04 \endcsname & 0 & 1 \\
\csname model_mn05 \endcsname & 0 & 1 \\
\csname model_mn06 \endcsname & 0 & 1 \\
\csname model_mn07 \endcsname & 7 & 1 \\
\csname model_mn08 \endcsname & 0 & 1 \\
\csname model_mn09 \endcsname & 10 & 0 \\
\csname model_mn010 \endcsname & 1 & 0 \\
\csname model_mn011 \endcsname & 10 & 0 \\
\csname model_mn012 \endcsname & 0 & 0 \\
\csname model_mn013 \endcsname & 6 & 3 \\
\csname model_mn014 \endcsname & 0 & 0 \\
\csname model_mn015 \endcsname & 10 & 0 \\
\csname model_mn016 \endcsname & 0 & 0 \\
\csname model_mn017 \endcsname & 1 & 1 \\
\csname model_mn018 \endcsname & 4 & 0 \\
\csname model_mn091 \endcsname & 0 & 0 \\
\csname model_mn092 \endcsname & 1 & 0 \\
\csname model_mn093 \endcsname & 2 & 0 \\
\csname model_mn094 \endcsname & 0 & 0 \\
\csname model_mn095 \endcsname & 1 & 0 \\
\csname model_imn01 \endcsname & 0 & 0 \\
\csname model_imn02 \endcsname & 1 & 0 \\
\csname model_imn03 \endcsname & 1 & 0 \\
\csname model_imn04 \endcsname & 0 & 0 \\
\csname model_mn028 \endcsname & 3 & 3 \\
\csname model_mn101 \endcsname & 0 & 0 \\
\csname model_mn102 \endcsname & 10 & 0 \\
\csname model_mn103 \endcsname & 9 & 1 \\
\csname model_mn104 \endcsname & 10 & 0 \\
\hline
\end{tabular}
\end{minipage}
\end{table}

If the \acp{IMBH} shall be left over in the \ac{UCD} after the dwarf galaxy was tidally stripped, they have to be inside the nucleus at that time. Otherwise, they would be removed with the rest of the disk.
We found that in our models with heavier \acp{IMBH}, some of the \acp{IMBH} were indeed trapped in the nucleus.
An example of this can be seen in Fig. \ref{fig_Core}, where we show the eight \acp{IMBH} of \csname model_mn03 \endcsname\ that were trapped in the nucleus.
It is clearly visible, that the \acp{IMBH} gather in the central region of the nucleus.
However, while those eight \acp{IMBH} are within 10 pc of the \ac{COM} not all \acp{IMBH} spiral in.
The remaining two \acp{IMBH} were at a distance of around 162 and 270 pc respectively with an initial distance of 176 pc for all \acp{IMBH} in this model.
These large radii would be due to the epicyclic motions of these \acp{IMBH} caused by the gravitational potential of their dwarf and interaction with other \acp{IMBH}.
We can also see that there are still plenty of stars in between the \acp{IMBH}.
In our simulation the stars belonging to the nucleus have a smaller softening length (0.5 pc) than the \acp{IMBH} (5.9 pc).
The large softening length prevents the \acp{IMBH} from getting closer to each other.
From our present simulation we cannot see whether the \acp{IMBH} would form an even tighter cluster.
We will discuss the possible further evolution of the central ``\ac{IMBH} cluster'' in section \ref{sec_Disc_GW}.

The final distances of the \acp{IMBH} to the \ac{COM} of all models are shown in Table \ref{table_FinalDist}.
As we can see for models with galaxy parameters similar to our fiducial one, central clusters only form in models with \ac{IMBH} masses of ${10^5~\rm M_\odot}$.
In our low mass models we have central clusters for models with $3 \times {10^4~\rm M_\odot}$ as well.
In our models with low \ac{IMBH} masses however only a single or no \ac{IMBH} reaches the \ac{COM}.

The five \acp{IMBH} to reach the \ac{COM} in \csname model_mn03 \endcsname\ first are shown in Fig. \ref{fig_centCluster} between ${\rm T}=0.816$ and ${0.832~\rm Gyr}$.
At this time the formation of the central cluster starts and it can be seen that the orbits of the \acp{IMBH} shrink rapidly due to dynamical friction against the stellar nucleus.
In the case of the first two \ac{IMBH} to reach the \ac{COM} they reach less than 10 pc at 0.832 Gyr.
We can also see the other \acp{IMBH} spiral in around them.
The exact behaviour of the \acp{IMBH} after forming a central cluster needs to be investigated in future simulations.

Although $M_{\rm bh} = {10^6~\rm M_\odot}$ is too big for IMBH thus would not be a reasonable in this parameter study, we analysed a model with $M_{\rm bh} = {10^6~\rm M_\odot}$ as an extreme test, to gain a better understanding of the importance of \ac{BH} masses in the orbital evolution of \acp{IMBH}.
The dwarf galaxy for this model has the same parameters as the fiducial model.
The results are discussed in Appendix \ref{App_LargeBH}.
Furthermore, the results for models with very massive ($M_{\rm dm} = { 3 \times 10^{10}~\rm M_\odot }$) and very low-mass (${3 \times 10^8~\rm M_\odot}$) dark matter halo are described in Appendix \ref{App_ExtGal}.

\begin{figure}
	\includegraphics{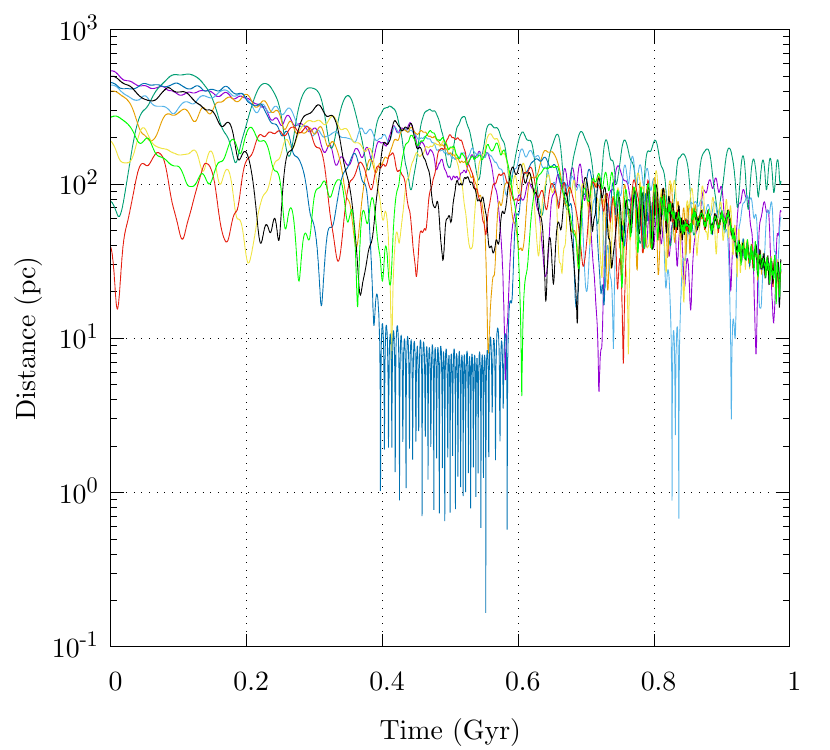}
	\caption{Time evolution of distances of different IMBHs in model \csname model_mn013 \endcsname\ (shown in different colours) to one specific IMBH in the model.
	It can clearly be seen that this IMBH formes a binary with another one visible in blue at around $T = {0.4~\rm Gyr}$.}
	\label{fig_Dist1_mn13}
\end{figure}

\begin{figure}
	\includegraphics{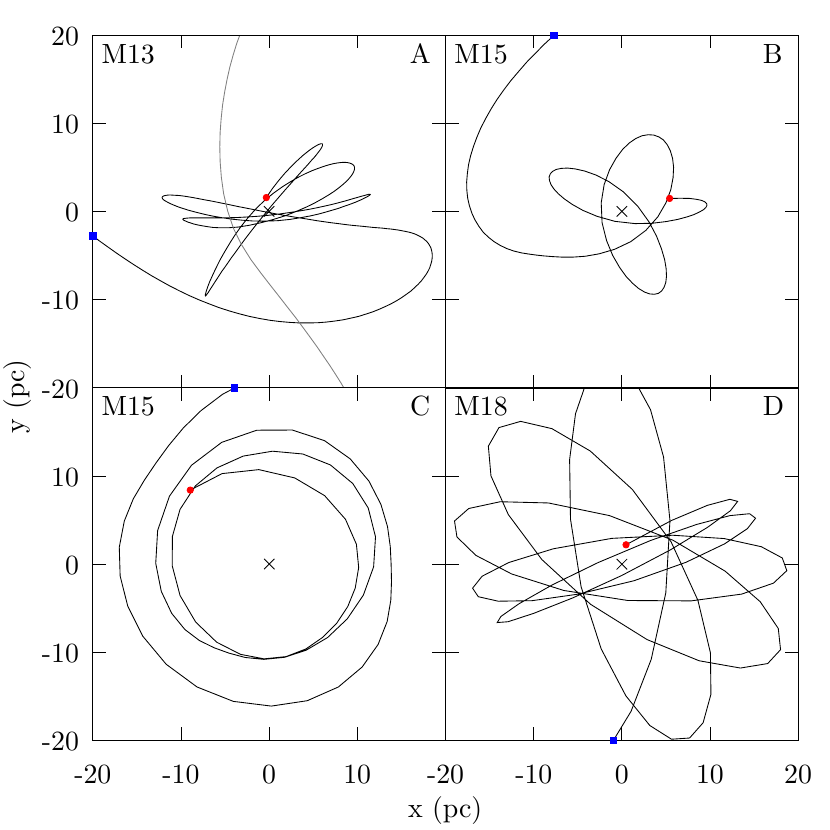}
	\caption{Four examples of IMBH binaries forming in our simulations.
	All of them are plotted relative to one of the IMBHs marked with a cross.
	The position of the IMBHs at the beginning of the time interval given below are shown with a blue square, at its end with a red dot.
	The model Ids are given at the top left of each panel.
	Panel A shows a binary forming in \csname model_mn013 \endcsname\ between $T={0.367~\rm Gyr}$ and $T={0.430~\rm Gyr}$.
	The orbit of a third IMBH that was interfering with the binary is plotted in grey.
	It enters from the top at $T={0.412~\rm Gyr}$ and leaves towards the bottom at $T={0.416~\rm Gyr}$.
	Panels B and C show binaries that formed in \csname model_mn015 \endcsname.
	They are shown between $T={0.056~\rm Gyr}$ and $T={0.080~\rm Gyr}$, and between $T={0.195~\rm Gyr}$ and $T={0.207~\rm Gyr}$ respectively.
	Finally, we show a binary that formed in \csname model_mn018 \endcsname\ in panel D.
	It formed between $T={0.550~\rm Gyr}$ and $T={0.564~\rm Gyr}$.
	It should be noted here that the binary orbits are quite diverse ranging from almost circular orbits to highly elongated ones.}
	\label{fig_Bin}
\end{figure}

\subsection{Binary formation}
\label{sec_Bin}

Another important aspect of this study is the formation of \ac{IMBH} binaries, which may lead to mergers emitting \acp{GW}.
It should be noted that our code cannot compute mergers or simulate the behaviour of close binaries accurately due to the large size resolution of the models of 5.9 pc and \ac{GW} physics not being included.

To spot binaries, we first plotted the distances of the \acp{IMBH} to each other.
An example of this can be seen in Fig. \ref{fig_Dist1_mn13}, which shows the distances of the \acp{IMBH} of \csname model_mn013 \endcsname\ to one of the binary forming \acp{IMBH}.
It can clearly be seen that one of the other \acp{IMBH} (visible in blue) stays within 10 pc to it between 0.4 and 0.6 Gyr.
The binary dissolves after two close encounters with other \ac{IMBH}.
After we identified the potential binary, we can have a look at its orbits.

The first 3 orbits of this binary after its formation are shown in Fig. \ref{fig_Bin} A.
As can be seen in this figure, their orbits are elongated (larger orbital eccentricities).
We have identified a third \ac{IMBH} (visible in yellow in Fig. \ref{fig_Dist1_mn13}), influencing the forming binary, which we plotted as a gray line.
This \ac{IMBH} seems responsible for the elongated orbit of the \ac{IMBH}.
At the time of the binary formation, the \acp{IMBH} were about 100 pc away from the galactic centre.
The \acp{IMBH} in Fig. \ref{fig_Bin} B are about 160 pc away from the \ac{COM} while forming an \ac{IMBH} binary and the ones in Figs. \ref{fig_Bin} C and \ref{fig_Bin} D are less than 20 pc away from the \ac{COM}.
The binaries in Figs. \ref{fig_Bin} B and \ref{fig_Bin} C belong to \csname model_mn015 \endcsname, while the one in Fig. \ref{fig_Bin} D formed in \csname model_mn018 \endcsname.
In addition to the binaries shown here, the two heaviest \acp{IMBH} of \csname model_mn093 \endcsname\ formed a binary when reaching the \ac{COM}.
We also had models where two \acp{IMBH} were part of a binary from the start.
\csname model_mn07 \endcsname, \csname model_mn09 \endcsname\ and \csname model_mn015 \endcsname\ had one initial binary each while two such binaries were present in \csname model_mn011 \endcsname.

From this we can conclude that, while binary formation is rare for our assumed initial \ac{IMBH} number before the \acp{IMBH} reach the \ac{COM}, binaries can form at any distance from it.
While the number of \acp{IMBH} could be higher in a real galaxy, in most of our models we put all \acp{IMBH} at a similar initial distance from the galactic \ac{COM}.
This does not necessarily need to be true in reality, making \acp{IMBH} encounters rarer.
The binary in Fig. \ref{fig_Bin} D stays bound for about 0.2 Gyr until other \acp{IMBH} join it and form a central cluster. 
Similarly, the other binaries we found stay bound for a few ${10^{-1}~\rm Gyr}$ until they dissolve, usually due to an encounter with another \ac{IMBH}. However, the majority of \acp{IMBH} reaches the \ac{COM} of the nucleus without becoming part of a binary prior to reaching the \ac{COM}.

Because of our large scale length, we cannot say if distinct binaries would form in the central \ac{IMBH} cluster or not.
However, if such binaries form, it is quite likely that those binaries would quickly harden due to encounters with nearby stars.
This is especially true in the presence of a nucleus, which would increase amount of stars near the \ac{COM}.
We will discuss the merger process of those \acp{IMBH} further in section \ref{sec_Disc_GW}.

\section{Discussion}
\label{sec_Disc}

\subsection{Can \ac{IMBH} mergers sufficiently explain \ac{MBH} formation?}
\label{sec_Disc_enoughBH}

In the last section we showed that \acp{IMBH} can spiral into the nucleus and then the get trapped there.
However, as we saw in section \ref{sec_DF} only heavy \acp{IMBH} spiral in.
For our fiducial model, which has a galaxy mass of ${1.036 \times 10^{10}~\rm M_\odot}$, their masses need to be at least around ${10^5~\rm M_\odot}$. 
The four central \acp{BH} we have observed in \acp{UCD} so far had masses between $3.5 \times 10^6$ \citep{10.1093/mnras/sty913} and $2.1 \times {10^7~\rm M_\odot}$ \citep{2014Natur.513..398S}.
Can \acp{BH} of this size be built from numerous \acp{IMBH} within the central regions of dwarf galaxies?

For this study we neglect the mass lost due to \ac{GW} radiation so that all of the mass from merging \acp{IMBH} adds to the final mass of the central \ac{MBH}.
Therefore, to create a light \ac{MBH} of ${10^6~\rm M_\odot}$, we need more than 10 of our heaviest \acp{IMBH}.
For the heaviest \ac{MBH} observed in an \ac{UCD} so far, that number rises to over 210.
We also want to note that not necessarily all \acp{IMBH} reach the \ac{COM}.
For example in \csname model_mn03 \endcsname\ only 80 per cent of the \acp{IMBH} are trapped in the nucleus.
Additionally, \cite{2019arXiv191207681R} found that \acp{IMBH} merging in \acp{GC} could experience strong recoil kicks with kick velocities of over ${10^3~\rm km~s^{-1}}$ due to asymmetric \ac{GW} emission.
This could be strong enough to eject the \acp{IMBH} from the nucleus again.
Therefore, the required number of \acp{IMBH} could be a lot higher than estimated here.

Quantifying the number of \acp{IMBH} expected to form in a galactic disk is difficult as \acp{IMBH} are hard to detect observationally.
As we saw heavy \acp{IMBH}, which we would require in our scenario, also spiral in within a relatively short time frame and therefore the window of opportunity to detect them is relatively short.
If the hypotheses for \ac{IMBH} formation we mentioned in the introduction are correct we could try to reduce the required number of \acp{IMBH} in a dwarf for the adopted \ac{MBH} formation scenario.
For example, if stellar nuclei are formed from merging \acp{GC} initially, then stellar nuclei might have a number of \acp{IMBH} already.
However, as discussed by previous studies of \acp{IMBH} in \acp{GC} \citep[e.g.][]{2019MNRAS.488.5340B}, the masses of \acp{IMBH} in massive galactic \acp{GC} can be rather small (${<10^4~\rm M_\odot}$).
So, this idea might not be so promising.

\subsection{\Ac{GW} radiation from binary \acp{IMBH}}
\label{sec_Disc_GW}

In our simulation we showed that especially heavy \acp{IMBH} spiral into the centre of the nucleus, where they form a cluster with less than 10 pc between individual \acp{IMBH}.
We also explained that we cannot resolve smaller distances between \acp{IMBH} due to our large softening distance.
What we would expect to happen after the binary formation would be the hardening of the binary due to the encounters with the surrounding stars.
If the binary reaches a certain distance energy loss through \ac{GW} generation becomes dominant and the binary merges emitting \acp{GW}.

When the \acp{IMBH} binary encounters stars it will harden by ejecting the lighter stars.
This is no problem as long as the \ac{IMBH} binary is still moving towards the \ac{COM}.
If the \acp{IMBH} are already in the centre of the galaxy, a loss cone will form around the binary.
The question becomes whether or not this loss cone can be refilled with stars quickly enough for the binary to harden sufficiently to reach the distance where \ac{GW} emission becomes efficient.
If this is not the case the binary evolution stalls at a distance of less than 1 pc.
This is known as the final parsec problem \citep{2003ApJ...596..860M}.
In our model, however, we saw that most \acp{IMBH} reach the \ac{COM} without becoming part of a binary.
Therefore, several \acp{IMBH} meet at the \ac{COM} of the dwarf galaxy, so that the interactions of the \acp{IMBH} among each other could lead to the quick formation of tight binaries and increase the number of mergers.
The same interactions could, on the other hand, also lead to ejections of \acp{IMBH} further increasing the required number of \acp{IMBH}.

However, could we detect the \acp{GW} emitted by our \ac{IMBH} mergers?
LISA's detection limits are shown in Fig. 2 from \cite{2019arXiv190804985J}.
According to this, LISA can detect \acp{BH} between $10^4$ and ${10^9 ~ \rm M_\odot}$ depending on the \ac{IMBH} binary's properties and its redshift.
The masses required by our model are within those boundaries.
Therefore, the mergers leading to a central \ac{MBH} in a \ac{UCD} or stellar nucleus should be detectable using LISA given the \acp{BH} are at a low enough redshift.
It remains to be investigated in the future how many \ac{IMBH} mergers are possible in \acp{UCD} and stellar nuclei of dwarfs for a fixed volume at low redshifts in order to estimate the detection rate of \ac{IMBH} merging in LISA.

\subsection{Future work}
\label{sec_Disc_FW}

One of the questions left open is: How many seed \acp{IMBH} can we expect in the disk of a dwarf galaxy and how massive are they?
\Ac{GW} detections by LISA might be able to shed some light on this issue by detecting \ac{IMBH} mergers.
However, as LISA can only detect \acp{IMBH} mergers and not single individual \acp{IMBH} in dwarfs, we need to look other effects like X-ray signals due to the accretion of cold gas on these single \ac{IMBH}.
Additionally, theoretical and numerical work quantifying the expected number of \acp{IMBH} in galactic disks is required.

A major part still to be investigated is the final stage of merging the \acp{IMBH}.
One possible approach would be to investigate only the final stage after the \acp{IMBH} were already trapped in the nucleus.
In this case the influence of the outer disk stars can be neglected.
Using this smaller system we have more resources to use smaller timesteps and a smaller softening length, which would allow us to observe the hardening of our binaries to get a better understanding of the time scales up to the actual merger.
In particular, this could tell us whether or not binaries can reach the distance at which \ac{GW} emission becomes important.
However, this distance is very small.
Even for very massive \acp{BH} (${10^8 ~ \rm M_\odot}$) the estimated distance at which \acp{GW} become important is at around ${10^{-2} ~ \rm pc}$ \citep{2016IAUS..312...92V}.
To simulate the last bit \ac{GW} physics would be required which means that we needed a completely different code.

Another question which would be answered in these more detailed simulations is what portion of the \acp{IMBH} contributes to the final \ac{MBH} and how many of them are ejected due to 3-body-interactions and recoil kicks.
\cite{2016PASA...33...36H} simulated a similar scenario for stellar mass \acp{BH} in \acp{GC}.
They found that the majority of \acp{BH} were ejected.
If similar results were found for \acp{IMBH} in dwarf galaxies this would heavily increase the number of \acp{IMBH} required by our model.
Additionally, it would be interesting to learn how these effects affect the dynamical evolution of the \ac{UCD}.

In our current work, we assumed that our system only consists of stars, dark matter and \acp{IMBH}.
However, for a more realistic model the influence of other components should be investigated as well.
Especially the accretion of gas could influence the result significantly.
While gas accretion would add some mass to the \acp{IMBH}, \ac{DF} is less efficient for gas than it is for stars due to the collisional nature of gas.
Therefore, the complete investigation we did here had to be repeated after adding gas, to investigate the effect on \ac{BH} dynamics as well.

\section{Conclusions}
\label{sec_conc}

We have investigated whether \ac{MBH} formation through the merging of \acp{IMBH} is possible in dwarf galaxies using Nbody simulations.
We have assumed that (i) each \ac{IMBH} is represented by a point-mass particle, (ii) \acp{IMBH} cannot grow through accretion of \ac{ISM} of their host dwarf galaxies, and (iii) \acp{IMBH} can gravitationally interact with other \acp{IMBH} and stars and dark matter of their hosts (no hydrodynamical interaction with \ac{ISM} owing to no inclusion of \ac{ISM}).
We have mainly investigated how dynamical friction of field stars of dwarf galaxies can influence the orbits of \acp{IMBH} within the dwarfs.
The principle results are as follows:

\begin{enumerate}[leftmargin=\parindent,align=left,labelwidth=\parindent,labelsep=0pt,label=\arabic*.]
\item Only the most massive \acp{IMBH} (${10^5 ~ \rm M_\odot}$ for large dwarf galaxies ($\approx 10^{10}~\rm M_\odot$) and $3 \times {10^4 ~ \rm M_\odot}$ for small ones ($\approx 10^{9}~\rm M_\odot$)) spiral in due to \ac{DF} of \acp{IMBH} against disk field stars within less than ${1~\rm Gyr}$.
	There is a positive correlation between dwarf mass and threshold mass for \acp{IMBH} being able to spiral into the dwarf's \ac{COM} quickly enough.
	However, in a galaxy with 3 times the mass of our fiducial model the threshold mass exceeds ${10^5~\rm M_\odot}$, while for a galaxy with a third of our low-mass model even ${10^3~\rm M_\odot}$ can spiral into the \ac{COM} within 1 Gyr.
	The masses of \acp{IMBH} and the dwarfs are the main factors that determine whether the \acp{IMBH} spiral in or not.
	\Ac{DF} is stronger for heavier \acp{IMBH} and in smaller dwarfs.
	The initial distance and the presence or absence of a nucleus have only little influence.
\item Binary \acp{IMBH} in dwarf galaxies can form both after the \acp{IMBH} reach the nucleus and before.
	However, the latter is quite rare.
	While we could observe binary formation in our simulations, we could not observe the hardening of the binaries due to the relatively large gravitational softening length used.
	As most binaries form inside the nucleus, simulating only the nucleus with a smaller softening length could give us further inside in the behaviour of \ac{IMBH} binaries.
	Therefore, future simulations using a different code, e.g. NBODY6, are required and for the final merger modelling energy loss due to \acp{GW} will be required as well.
\item We expect that merging of massive \acp{IMBH} with $M_{\rm bh} \sim {10^5~\rm M_\odot}$ can occur mostly in the central regions of dwarfs.
    Given theese high masses, we expect the merger to emit \acp{GW} that could be detected by LISA if they are within dwarfs at lower redshifts.
\item At present we cannot say how many \acp{IMBH} should be expected in the disk of a dwarf galaxy.
	Our model requires at least 10 \acp{IMBH} for a small \ac{MBH} (${10^6 ~ \rm M_\odot}$) to form.
	The required number is even larger if we take into account that not all \acp{IMBH} necessarily reach the centre and that mass is lost due to \ac{GW} and possibly ejections.
	The number of \acp{IMBH} required for large \acp{MBH} to form is quite large ($>200$ for the heaviest \ac{MBH} found in an \ac{UCD} thus far).
	It is unknown if such a high number of \acp{IMBH} can exist in the disk of a dwarf galaxy.
	Therefore, future simulations should investigate the role of gas accretion in \ac{MBH} formation as well.
\end{enumerate}

From the present work, we can conclude that it is possible that the \acp{MBH} observed in \acp{UCD} formed through \acp{IMBH} mergers, though a high enough number of \acp{IMBH} initially in dwarfs is required.
However, more research is required to confine the possible number and masses \acp{IMBH} in the disk of a dwarf galaxy.
Additional more detailed investigations of the hardening of \ac{IMBH} binaries and their mergers are needed.
Investigating other processes contributing to \ac{BH} growth, such as gas accretion, would contribute to completing the picture as well.

\section*{Acknowledgements}

We are  grateful to the referee  for  constructive and useful comments that improved this paper.

\bibliographystyle{mn2e}
\bibliography{imb}

\FloatBarrier

\appendix
\section{Simulations with larger BHs}
\label{App_LargeBH}

\begin{figure}
	\includegraphics{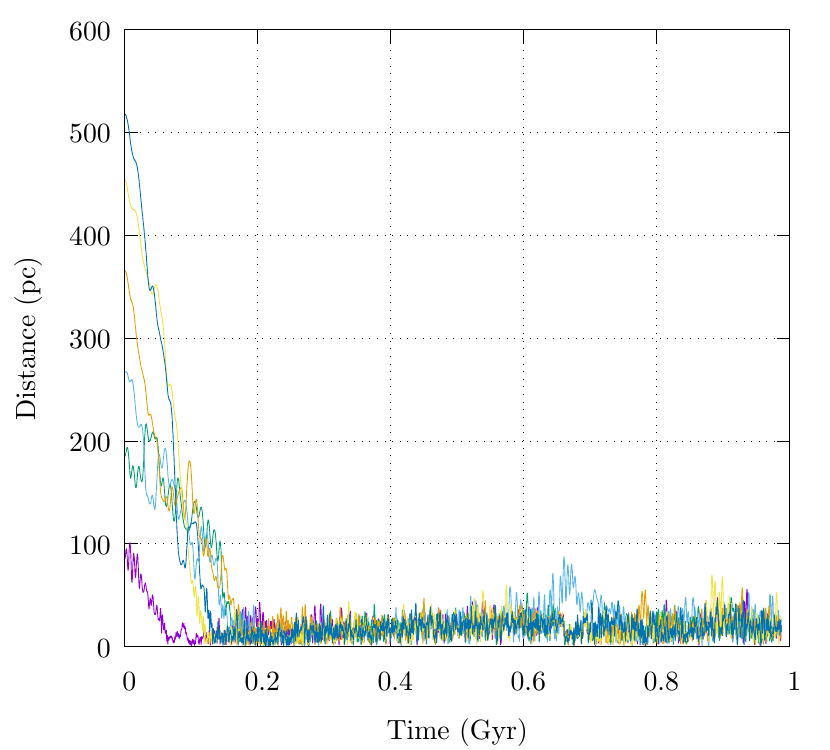}
	\caption{The six IMBHs of \csname model_mn028 \endcsname.
	This models parameter's are the same as for the fiducial model, which means that the dwarf's mass is at ${1.036 \times 10^{10}~\rm M_\odot}$.
	However, the IMBH in this model have a mass of ${10^6~\rm M_\odot}$.
	Different colours denote different BHs.}
	\label{fig_Dist_MBH}
\end{figure}

In addition to testing the effects of \ac{DF} on \acp{IMBH}, we tested what happens if we have ${10^6~\rm M_\odot}$ \acp{BH} from the start in model \csname model_mn028 \endcsname.
The evolution of the \ac{MBH} orbits can be seen in Fig. \ref{fig_Dist_MBH} in the appendix.
The \ac{BH} in this model reach the galaxy's \ac{COM} after less then 0.2 Gyr and, therefore, a lot faster than for example the \acp{BH} in \csname model_mn03 \endcsname, which are a factor of 10 lighter, while the galaxy's parameter stay the same.
This is to be expected, because the more massive \acp{BH} have a larger sphere of influence, therefore increasing the deceleration the \acp{BH} experience.
In Table \ref{table_FinalDist} we can see, that only 3 of the six \ac{MBH} end up within the inner 20 pc of the nucleus.
The other 3 are within the central 50 pc.
This is due to the highly eccentric orbits the \acp{MBH} have due to their interactions with one another.
This leads to a total \ac{BH} mass of ${6 \times 10^6~\rm M_\odot}$ in the central cluster, which is within the range of observed \acp{MBH} in \acp{UCD} ($3.5 \times 10^6$ to $2.1 \times 10^7~\rm M_\odot$).
No evidence of binaries forming before the \acp{MBH} reach the galaxies \ac{COM}.
Because of our small sample we cannot derive a general conclusion from this.
We can, however, conclude from the quick descent of the \acp{MBH} into the dwarf's \ac{COM} that such an event would be highly improbable as there is only little time for those binaries to form.

\section{Simulations with different galaxy masses}
\label{App_ExtGal}

\begin{figure}
	\includegraphics{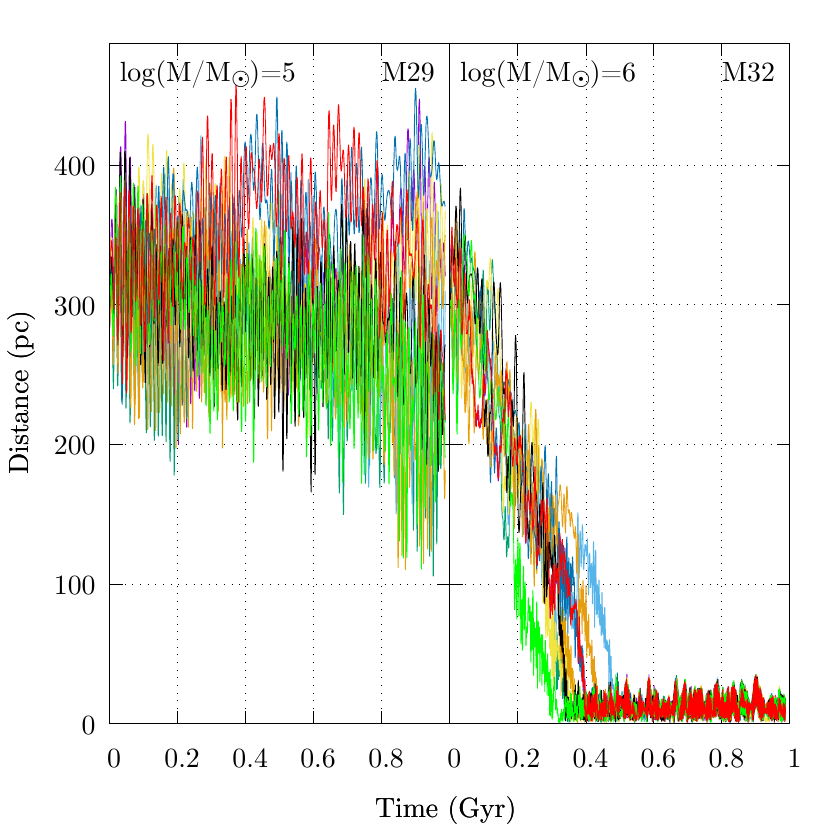}
	\caption{Two different models for a heavy dwarf.
	The model ID can be seen in the top right, while the BH mass is shown in the top left of each panel.
	Different colours denote different IMBHs.
	With a mass of ${3.108 \times 10 ^{10}~\rm M_\odot}$ this galaxy's mass is 3 per cent of that of the Milky Way and 3 times that of our fiducial model.}
	\label{fig_Dist_bigGal}
\end{figure}

\begin{figure}
	\includegraphics{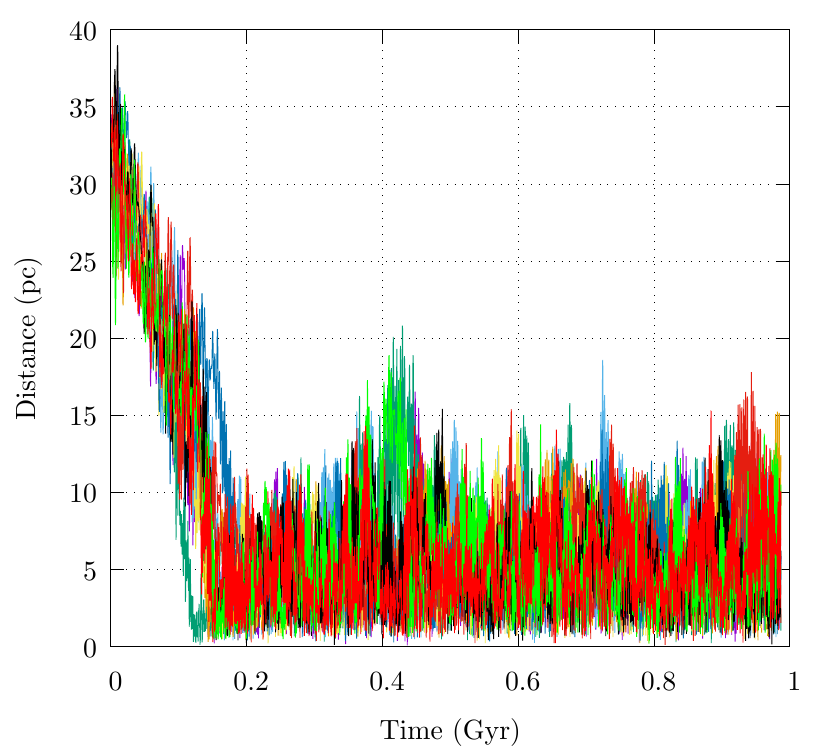}
	\caption{10 $10^4~\rm M_\odot$ IMBHs in a smaller galaxy in model \csname model_mn102 \endcsname.
	Different colours denote different IMBHs.
	This galaxy's mass is ${3.1 \times 10^8~\rm M_\odot}$, less than a third of that of our low-mass models.}
	\label{fig_Dist_smaGal}
\end{figure}

The results of our models with the heaviest dwarf, \csname model_mn101 \endcsname\ and \csname model_mn104 \endcsname, can be seen in Fig. \ref{fig_Dist_bigGal}.
While the distances of the \acp{IMBH} to the dwarf's \ac{COM} fluctuate rapidly in model \csname model_mn101 \endcsname, none of the \acp{IMBH} gets to a distance closer than 100 pc.
No general trend is visible.
This can be explained through the high velocities of the stars that lead to them only having short encounters with the \acp{IMBH} and therefore a low \ac{DF} deceleration.
Because of this the threshold mass for \ac{BH} of ${10^5~\rm M_\odot}$ we found for our fiducial model does not apply here.
As we can see looking at model \csname model_mn104 \endcsname\ ${10^6~\rm M_\odot}$ \acp{BH} do spiral into the dwarf's \ac{COM}.
This means that the threshold mass is shifted to higher values.

The opposite effect can be seen for models \csname model_mn102 \endcsname\ and \csname model_mn103 \endcsname.
As a result of their low velocities, \ac{DF} is very efficient and we can see the \acp{IMBH} move towards the dwarf's \ac{COM} quickly.
An example of this can be seen in Fig. \ref{fig_Dist_smaGal}, where ten ${10^4~\rm M_\odot}$ \acp{IMBH} move to an orbit with less than 20 pc in less than 0.2 Gyr.
From Table \ref{table_FinalDist} we can see, that even small ${10^3~\rm M_\odot}$ \acp{IMBH} reach the \ac{COM} in less than one Gyr.
This continues the trend we already observed, namely that the threshold mass for \acp{IMBH} to be able to reach the \ac{COM} within 1 Gyr is lower in lighter dwarfs and higher in heavier dwarfs.
While there is an initial binary in both \csname model_mn102 \endcsname\ and \csname model_mn103 \endcsname, no binaries form during the evolution of any of the four models discussed here before the \acp{IMBH} reach the dwarf's \ac{COM}.

\end{document}